\documentclass[conference]{IEEEtran}
\usepackage{cite}

\ifCLASSINFOpdf
  \usepackage[pdftex]{graphicx}

\else

\fi
\usepackage{amsmath}
\usepackage{url}

\usepackage[shortcuts,acronym]{glossaries}
\usepackage{graphicx}
\usepackage{multirow}
\usepackage{booktabs}
\usepackage{diagbox}
\usepackage{colortbl}
\usepackage[table]{xcolor}

\newacronym{nft}{NFT}{Non-fungible token}
\newacronym{dex}{DEX}{Decentralized Exchange}
\newacronym{cex}{CEX}{Centralized Exchange}
\newacronym{pow}{PoW}{Proof of Work}
\newacronym{utxo}{UTXO}{Unspent Transaction Output}
\newacronym{ringct}{RingCT}{Ring Confidential Transactions}

\newacronym{tpr}{TPR}{True Positive Rate}
\RequirePackage{xspace,xcolor}

\usepackage{xspace}

\newcommand{\nMoneroMainnetStartDate}{2014-04-18\xspace}
\newcommand{\nMoneroMainnetEndDate}{2023-10-31\xspace}

\newcommand{\nACKJMoneroTestnetStartDate}{2022-01-20\xspace}
\newcommand{\nACKJMoneroTestnetEndDate}{2022-02-23\xspace}
\newcommand{\nACKJMoneroHeuristicPrecision}{\ensuremath{\approx 99.998\%}\xspace}

\newcommand{\nACKJMoneroTestnetTransactions}{\ensuremath{760,588}\xspace}
\newcommand{\nACKJMoneroTestnetKeyImages}{\ensuremath{1,334,693}\xspace}
\newcommand{\nACKJMoneroTestnetKeyImagesSingleTenBlock}{\ensuremath{92,954}\xspace}

\newcommand{\nDifferByOneTransactions}{\ensuremath{58,429}\xspace}
\newcommand{\nDifferByOneStartDate}{2014-04-21\xspace}
\newcommand{\nDifferByOneEndDate}{2023-10-31\xspace}

\newcommand{\nMordinalsMinted}{\ensuremath{43,099}\xspace}
\newcommand{\nMordinalDecoys}{\ensuremath{474,442}\xspace}
\newcommand{\nMordinalFalsePositives}{\ensuremath{21}\xspace}
\newcommand{\nMordinalTruePositives}{\ensuremath{9,934}\xspace}
\newcommand{\nMordinalPrecision}{\ensuremath{99.79\%}\xspace}

\newcommand{\nChainReactionIncrementalSpentOutputs}{\ensuremath{40,005}\xspace}
\newcommand{\nChainReactionIncrementalKnownDecoys}{\ensuremath{61,928}\xspace}

\newcommand{\nSelfCollisionsZeroMixCascade}{\ensuremath{0\%}\xspace}
\newcommand{\nSelfCollisionsDifferByOne}{\ensuremath{< 0.01\%}\xspace}
\newcommand{\nSelfCollisionsTenBlock}{\ensuremath{0.02\%}\xspace}
\newcommand{\nSelfCollisionsCoinbase}{\ensuremath{0\%}\xspace}
\newcommand{\nSelfCollisionsMordinals}{\ensuremath{0\%}\xspace}
\newcommand{\nSelfCollisionsPPool}{\ensuremath{0.04\%}\xspace}

\newcommand{\nCollisionsMordinalsZeroMixCascade}{\ensuremath{0\%}\xspace}
\newcommand{\nCollisionsMordinalsDifferByOne}{\ensuremath{0.12\%}\xspace}
\newcommand{\nCollisionsMordinalsTenBlock}{\ensuremath{0.21\%}\xspace}
\newcommand{\nCollisionsMordinalsCoinbase}{\ensuremath{0\%}\xspace}

\newcommand{\nCollisionsCoinbaseZeroMixCascade}{\ensuremath{0.04\%}\xspace}
\newcommand{\nCollisionsCoinbaseDifferByOne}{\ensuremath{0\%}\xspace}
\newcommand{\nCollisionsCoinbaseTenBlock}{\ensuremath{0\%}\xspace}

\newcommand{\nCollisionsTenBlockZeroMixCascade}{\ensuremath{0\%}\xspace}
\newcommand{\nCollisionsTenBlockDifferByOne}{\ensuremath{0.39\%}\xspace}

\newcommand{\nCollisionsDifferByOneZeroMixCascade}{\ensuremath{10.02\%}\xspace}

\newcommand{\nAgreementsMordinalsZeroMixCascade}{\ensuremath{0\%}\xspace}
\newcommand{\nAgreementsMordinalsDifferByOne}{\ensuremath{0.36\%}\xspace}
\newcommand{\nAgreementsMordinalsTenBlock}{\ensuremath{2.09\%}\xspace}
\newcommand{\nAgreementsMordinalsCoinbase}{\ensuremath{0\%}\xspace}

\newcommand{\nAgreementsCoinbaseZeroMixCascade}{\ensuremath{0.04\%}\xspace}
\newcommand{\nAgreementsCoinbaseDifferByOne}{\ensuremath{< 0.01\%}\xspace}
\newcommand{\nAgreementsCoinbaseTenBlock}{\ensuremath{2.14\%}\xspace}

\newcommand{\nAgreementsTenBlockZeroMixCascade}{\ensuremath{< 0.01\%}\xspace}
\newcommand{\nAgreementsTenBlockDifferByOne}{\ensuremath{0.88\%}\xspace}

\newcommand{\nAgreementsDifferByOneZeroMixCascade}{\ensuremath{8.76\%}\xspace}

\newcommand{\nCollisionsPPoolZeroMixCascade}{\ensuremath{0\%}\xspace}
\newcommand{\nCollisionsPPoolDifferByOne}{\ensuremath{0\%}\xspace}
\newcommand{\nCollisionsPPoolTenBlock}{\ensuremath{5.41\%}\xspace}
\newcommand{\nCollisionsPPoolCoinbase}{\ensuremath{10.19\%}\xspace}
\newcommand{\nCollisionsPPoolMordinals}{\ensuremath{0\%}\xspace}

\newcommand{\nAgreementsPPoolZeroMixCascade}{\ensuremath{0\%}\xspace}
\newcommand{\nAgreementsPPoolDifferByOne}{\ensuremath{0.07\%}\xspace}
\newcommand{\nAgreementsPPoolTenBlock}{\ensuremath{1.84\%}\xspace}
\newcommand{\nAgreementsPPoolCoinbase}{\ensuremath{36.42\%}\xspace}
\newcommand{\nAgreementsPPoolMordinals}{\ensuremath{0.09\%}\xspace}

\newcommand{\nOverlapChainReaction}{}
\newcommand{\nOverlapChainReactionDifferByOne}{}
\newcommand{\nOverlapChainReactionTenBlock}{}
\newcommand{\nOverlapChainReactionCoinbase}{}
\newcommand{\nOverlapChainReactionMordinals}{}
\newcommand{\nOverlapChainReactionPPool}{}

\newcommand{\nOverlapDifferByOne}{}
\newcommand{\nOverlapDifferByOneTenBlock}{}
\newcommand{\nOverlapDifferByOneCoinbase}{}
\newcommand{\nOverlapDifferByOneMordinals}{}
\newcommand{\nOverlapDifferByOnePPool}{}

\newcommand{\nOverlapTenBlock}{}
\newcommand{\nOverlapTenBlockCoinbase}{}
\newcommand{\nOverlapTenBlockMordinals}{}
\newcommand{\nOverlapTenBlockPPool}{}

\newcommand{\nOverlapCoinbase}{}
\newcommand{\nOverlapCoinbaseMordinals}{}
\newcommand{\nOverlapCoinbasePPool}{}

\newcommand{\nOverlapMordinals}{}
\newcommand{\nOverlapMordinalsPPool}{}

\newcommand{\nOverlapPPool}{}

\begin{document}
\title{Monero Traceability Heuristics: Wallet Application Bugs and the Mordinal-P2Pool Perspective}

\author{
\IEEEauthorblockN{Nada Hammad}
\IEEEauthorblockA{TRM Labs\\
San Francisco, USA\\
nada@trmlabs.com}
\and
\IEEEauthorblockN{Friedhelm Victor}
\IEEEauthorblockA{TRM Labs\\
San Francisco, USA\\
friedhelm@trmlabs.com}
}

\maketitle

\begin{abstract}
Privacy-focused cryptoassets like Monero are intentionally difficult to trace. Over the years, several traceability heuristics have been proposed, most of which have been rendered ineffective with subsequent protocol upgrades. Between 2019 and 2023, Monero wallet application bugs ``Differ By One'' and ``10 Block Decoy Bug'' have been observed and identified and discussed in the Monero community. In addition, a decentralized mining pool named P2Pool has proliferated, and a controversial UTXO NFT imitation known as Mordinals has been tried for Monero.
In this paper, we systematically describe the traceability heuristics that have emerged from these developments, and evaluate their quality based on ground truth, and through pairwise comparisons. We also explore the temporal perspective, and show which of these heuristics have been applicable over the past years, what fraction of decoys could be eliminated and what the remaining effective ring size is.
Our findings illustrate that most of the heuristics have a high precision, that the ``10 Block Decoy Bug`` and the Coinbase decoy identification heuristics have had the most impact between 2019 and 2023, and that the former could be used to evaluate future heuristics, if they are also applicable during that time frame.

\end{abstract}

\IEEEpeerreviewmaketitle

\section{Introduction}

In the evolving landscape of digital currencies, Monero has emerged as a significant player, renowned for its strong emphasis on privacy. Originating from the CryptoNote protocol~\cite{van2013cryptonote}, Monero represents a key advancement in blockchain technology by employing stealth addresses, hiding transaction amounts using \gls{ringct}~\cite{noether2016ring} and obfuscating its underlying transaction graph through the use of ring signatures for transaction inputs.

As of November 2023, with a market capitalization of approximately 3 billion USD, Monero stands as the foremost privacy-focused cryptocurrency. This prominence can largely be attributed to its enduring commitment to privacy-by-default features, which notably surpass those offered by more mainstream blockchains such as Bitcoin and Ethereum. Unlike these transparent blockchains, where transactions are prone to traceability through heuristic methods that exploit protocol specifics or typical user behaviors~\cite{deuber2022sok,ghesmati2022sok,victor2020address}, Monero has consistently evolved to address its vulnerabilities.

However, despite its advancements, Monero has not been impervious to challenges. Over the years, various Monero-specific heuristics have been employed to probe its privacy features, leading to the identification of early protocol weaknesses~\cite{kumar2017traceability,moser2018empirical,hinteregger2019short,yu2019new}.
These vulnerabilities have been largely mitigated by protocol upgrades, wallet improvements, and ongoing community education. Yet, new concerns have arisen.

Recent discoveries of software bugs and the proposition of projects misaligned with Monero's privacy efforts (likely unintentionally), highlight ongoing threats to the currency's privacy framework. Several of these issues have previously been discussed in the Monero community, disclosed through Github issues and discussed on Reddit. But in this paper, we aim to formalize and assess the impact of these developments by studying the new heuristics that are applicable as a consequence.

Our contributions in this study are manifold. We describe the '10 Block Decoy Bug', 'Differ-by-one', Coinbase and Mordinal heuristics, as well as a P2Pool specific output merging heuristic based on publicly known miner payouts. We provide a systematic, comparative and combined evaluation, which constitutes the main contribution of this work. We present a comprehensive analysis, measuring the impact of each heuristic's applicability over time, as well as the effective ring size of Monero's transaction inputs up until October of 2023. This combined approach offers a nuanced understanding of the current state of privacy in the Monero ecosystem.

\section{Background and Related Work}
Monero is based on the cryptonote protocol~\cite{van2013cryptonote}. Its three most prominent privacy enhancing technologies are ring signatures, stealth addresses and \gls{ringct}~\cite{noether2016ring}. In contrast to a transparent \gls{utxo} blockchain design like in Bitcoin~\cite{nakamoto2008bitcoin}, each transaction input is a ring signature, where only one ring member is the truly spent input, and the others are decoys. As of 2023, the mandatory ring size in Monero is 16, meaning 15 ring members are decoys. Stealth addresses are one-time addresses generated by the sender of a transaction, derived from a supplied public address. \gls{ringct} hides the transaction amounts using bulletproofs~\cite{bunz2018bulletproofs}.

An input ring $R = \{pk_1,...,pk_n\}$ consists of a set of referenced transaction outputs, identified by their public key $pk$. Only one $pk \in R$ is the truly spent output, also known as the \textit{true spend}. As output $pk$ are used as inputs, one can refer to a $pk$ as an \textit{enote} in the general case. The input ring size is equal to the number of ring members it contains. If some input ring members can be identified as decoys, we refer to the remaining ring size as the \textit{effective ring size}.

This does not necessarily reveal the true spend, but reduces the anonymity set. An input ring of a monero transaction is said to be fully traceable, if the ring member that was truly spent is identified, and thereby all other ring members are marked as decoys.

\begin{figure}[h]
\centering
\includegraphics[width=\linewidth]{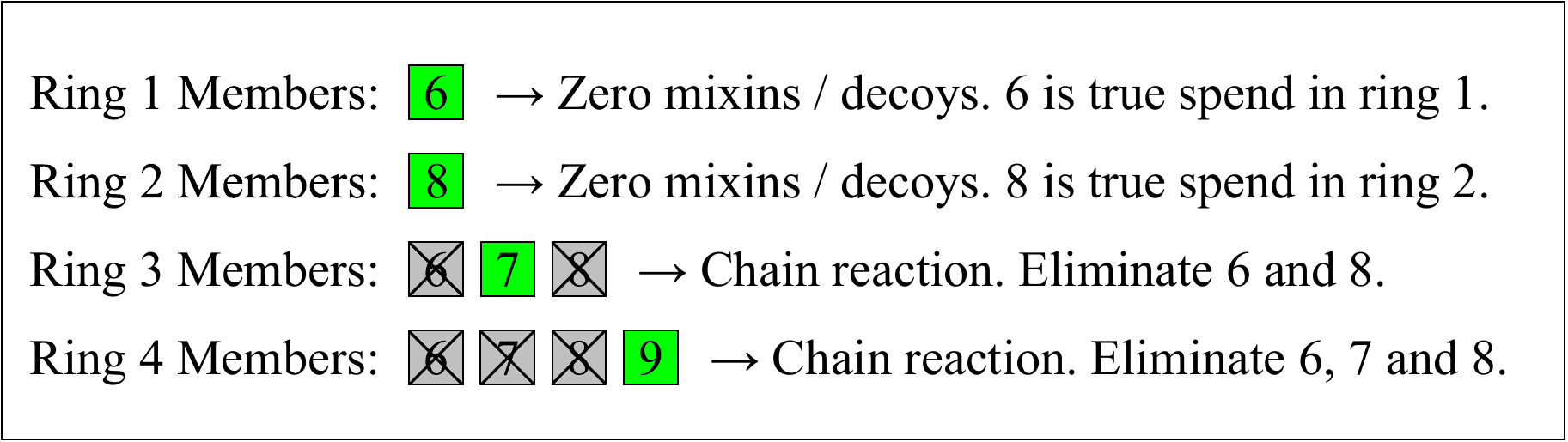}
\caption{Illustration of the zero mixins and chain reaction heuristics. Input rings 1 and 2 only have a single member, which means it must be the true spend. As enotes 6 and 8 are known to have been spent, it follows that enote 7 is the true spend of ring 3. The same approach works to identify output 9 as the true spend of ring 4, and is known as the chain reaction heuristic.}
\label{fig:zero-mixins}
\end{figure}

Early works on traceability heuristics for Monero have primarily focused on true spend identification heuristics.
In the two earliest and well known analyses by M\"oser et al.~\cite{moser2018empirical} and Kumar et al.~\cite{kumar2017traceability} traceability heuristics exploiting zero-mixins, chain reaction (also known as cascade), output-merging and guess newest have been explored. The zero-mixin heuristic identifies true spends in input rings that have only a single ring member, meaning no decoys -- formerly referred to as mixins. Such zero-mixin transactions were frequent until April 2016, but newer transactions are forced to have a minimum ring size. The chain reaction heuristic eliminates input ring members that are known to have been spent elsewhere. If all but one can be eliminated, the true spend has been identified. See Figure~\ref{fig:zero-mixins} for an example.
\vspace{0.1cm}

In 2019, Hinteregger and Haslhofer~\cite{hinteregger2019short} have proposed the intersection removal heuristic, exploiting the observability of the same key image associated with a spent output on forked Monero blockchains with a shared history. Here, the true spend may be identified in the intersection of the sets of input ring members. They have also shown that the guess newest heuristic no longer works due to an upgrade of Monero's decoy selection algorithm.
At the same time, Yu et al.~\cite{yu2019new} have proposed the closed set heuristic, which does not directly identify the ring in which a certain output has been spent, but can make a claim that the output must have been spent by a given point in time. It is therefore primarily a decoy identification heuristic.

Wijaya et al. have proposed a restricted version of the output merging heuristic applied to mining pool outputs~\cite{wijaya2021transparency}, but did not evaluate with ground truth~\cite{wijaya2021transparency}. They also studied the unforkability of Monero~\cite{wijaya2019unforkability}, as it allows for the intersection removal heuristic.

A recent work by Vijayakumaran~\cite{vijayakumaran2023analysis} showcased how the Dulmage-Mendelsohn Decomposition~\cite{dulmagemendelsohn1958} can be used to infer the true spends of closed sets in addition to the zero-mixin and cascade attacks, providing a polynomial-time implementation of the closed set heuristic.

Aeeneh et al.~\cite{aeeneh2021new} proposed methods to assess the probability of a ring member being the true spend: when a ring member appears only once and when the age distribution of ring members deviates from the expected true spend distribution. However, they did not empirically evaluate their work.

Finally, and highly relevant to this paper, anonymous user Rucknium conducted a preliminary privacy analysis of Mordinals and P2Pool outputs in an informal Reddit post\cite{rucknium2023reddit}.

\section{Definitions and Datasets}
Before we begin with our analysis, we define the general types of heuristics that we analyze, and describe our evaluation approach as well as the datasets used.
\subsection{Definitions}
Let $TX$ denote the set of all transactions stored on the Monero blockchain. Each transaction $tx \in TX$ creates outputs, identified by a public key $pk$, and a global output index $i$. An output is controlled by a set of public and private keys, typically belonging to a user or a service. A transaction input is an input ring $R = \{pk_1,...,pk_n\}$, where exactly one $pk_t \in R$ is the true spend, and the other ring members are decoys.

We fundamentally differentiate Monero traceability heuristics that are intended to identify true spends from those that identify decoys:
\begin{itemize}

    \item \textbf{True Spend Identification Heuristic}: For an input ring \( R_1 \), identifies the likely true spend \( pk_t \) in \( R_1 = \{pk_1,...,pk_n\} \). All other \( pk \in R_1 \) where \( pk \neq pk_t \) are decoys. If \( pk_t \) appears in any other input ring \( R_x \), it is marked as a decoy. This is based on the unique expendability of an enote in a single input ring. In summary, heuristics of this category identify true spends and, as a consequence, decoys, which may also appear in other input rings, and at a future point in time.
    
    \item \textbf{Decoy Identification Heuristic}: For an input ring \( R_1 \) determines a subset \( D_1 = \{pk_{d1}, ..., pk_{dm}\} \subseteq R_1 \) of public keys as decoys within \( R_1 \). The subset \( D \) can contain multiple enotes, based on specific criteria or be empty. In summary, heuristics of this category only identify decoys within input rings it is applied to.
    
\end{itemize}

\subsection{Evaluation Strategy}

We evaluate each heuristic against the well known zero-mixin and chain reaction heuristics results (also known as zero-decoy and cascade). While these heuristics have been largely ineffective since 2018~\cite{hinteregger2019short}, they still constitute one of the main sets of labels that can be treated as ground truth, as they have no false positives.
We will also evaluate most heuristics using the 10 block decoy bug heuristic that we will introduce in the first subsection~\ref{subsec:10_block_decoy_bug}, as it is a high confidence heuristic with results until 2023.

For true spend identification heuristics, we provide information on correctly identified true spends. For all heuristics, we evaluate on the basis of labeled ring members, for which we measure true positives (TP), false positives (FP) and precision $P$, with:
\[P = \frac{TP}{TP + FP}\]

We also measure collision rates between heuristics, and with themselves.
A collision occurs for example when heuristic $H_1$ claims that the same public key appearing in two different input rings ($pk_x \in R_1,R_2$), is the true spend. It can only be the true spend in one of them. A collision also occurs if a $H_1$ considers two ring members ($pk_x \in R_1, pk_y \in R_1$) to be the true spend of a single input ring $R_1$. There can only be one true spend per ring. If $C$ is the number of conflicting labels heuristic $H_1$ produces with itself, and the total number of ring members labeled by $H_1$ is $N$, we calculate the \textit{self collision rate} $SCR$ with:
\[SCR = \frac{C}{N}\]

\subsection{Datasets}

We have used three different datasets:
\begin{itemize}
    \item Monero mainnet blockchain transactions from the blockchain's inception on \nMoneroMainnetStartDate until \nMoneroMainnetEndDate for all analyses.
    \item Monero testnet transactions performed by ACK-J~\cite{ackj2022lotr} between \nACKJMoneroTestnetStartDate and \nACKJMoneroTestnetEndDate consisting of \nACKJMoneroTestnetTransactions transactions and \nACKJMoneroTestnetKeyImages input rings for which the true spend is known. This dataset consists of a large number of quickly spent outputs (20 minute interval target), making it suitable to study the 10 Block Decoy Bug (c.f. Section~\ref{subsec:10_block_decoy_bug}).
    \item 31,759 P2Pool mining pool payout transactions with 2,298,927 individual payouts scraped from p2pool.observer\footnote{\url{https://p2pool.observer}}. We use this dataset to study the p2pool output merging heuristic in Section~\ref{subsec:p2pool_output_merging}.
\end{itemize}

\section{Analysis}
We now define and empirically cross-evaluate 6 heuristics.

\subsection{10 Block Decoy Bug}
\label{subsec:10_block_decoy_bug}

On May 23, 2023, a significant vulnerability was disclosed within a wallet library, specifically in the `wallet2` component, as documented in the Monero project's GitHub issue tracker\footnote{\url{https://github.com/monero-project/monero/issues/8872}}. This bug, affecting the decoy selection algorithm, was identified in a library that forms the backbone of several widely-used Monero wallet applications, including Monero Wallet GUI/CLI, Feather Wallet, Cake Wallet, and Monerujo\footnote{\url{https://github.com/monero-project/research-lab/issues/99}}.
The flaw was an off-by-one error in the decoy selection process, leading to an inability to select decoys that were exactly 10 blocks old. This specific age of 10 blocks is particularly noteworthy because it represents the unlock time for outputs in Monero, marking the earliest point at which a user is able to spend a received output.

It was first communicated that this vulnerability was present across multiple versions of the wallet, from version v0.14.1.0 to v0.18.2.1. Introduced\footnote{\url{https://github.com/monero-project/monero/pull/5389}} initially on Apr 18, 2019, and fixed on April 10, 2023, which shows that this critical issue went undetected in the Monero ecosystem for almost 4 years. It was later commented that the issue may have been present since version v0.13.0.0, corresponding to October 11, 2018. To be on the safe side, and also account for users upgrading at later points in time, we assume the later date.
As a consequence of the vulnerability, we can define a time constrained heuristic that we refer to as the \textit{10 Block Decoy Bug Heuristic}, which is a true-spend identification heuristic that we define as follows:

\begin{figure}[!t]
\centering
\includegraphics[width=\linewidth]{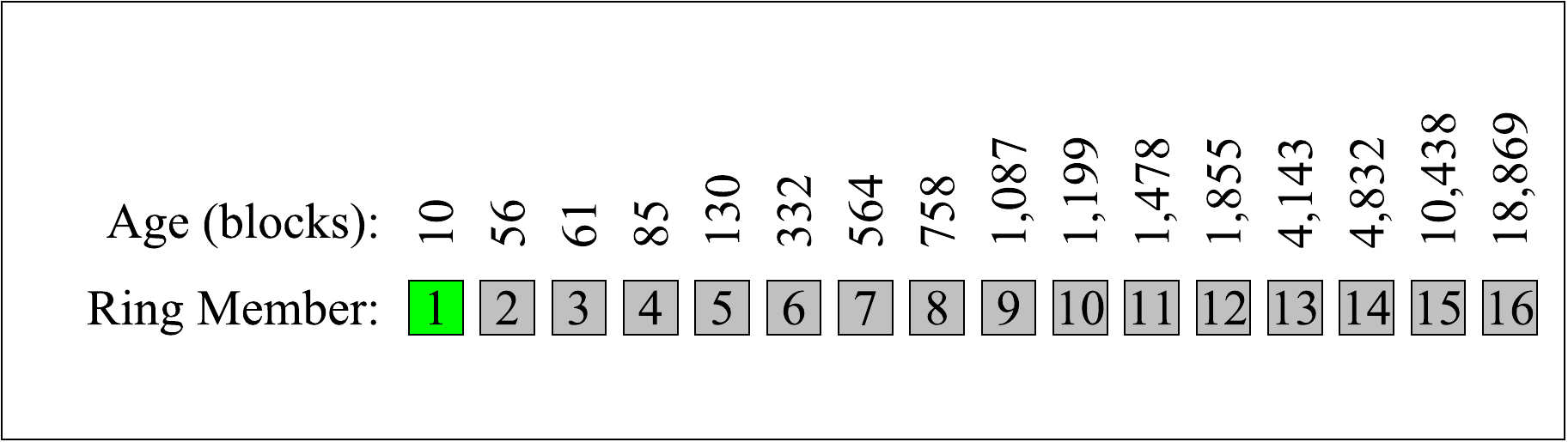}
\caption{Illustration of the 10-Block-Old Decoy Bug Heuristic: if there exists exactly one ring member that is 10 blocks old and the input ring has been created between October 11, 2018 and April 10, 2023, it is very likely the true spend (highlighted in green).}
\label{fig:10_block_decoy_bug}
\end{figure}

\noindent\textbf{Heuristic Definition.} Define \( age(pk, R) \) as a function that determines the age of a ring member \( pk \in R \) in terms of blocks. If there exists exactly one \( pk \) in \( R \) such that \( age(pk, R) = 10 \), then this \( pk \) is highly likely to be the true spend for input rings created between April 18, 2019 and April 10, 2023. An example is visualized in Figure~\ref{fig:10_block_decoy_bug}.

\noindent\textbf{Evaluation.} To validate the heuristic, we take two approaches:
\begin{enumerate}
    \item We evaluate the heuristic on the Monero Mainnet against ground truth obtained from the zero mixin and chain reaction heuristics. The heuristic identifies $1,365,175$ spent outputs and consequently $27,308,717$ decoys. In comparison with the zero mixin + chain reaction heuristic, there is an overlap of $3$ correctly identified true spends, and a total of $209$ true positives in terms of ring members labeled correctly. There are no false positives, leading to a precision of $100$\%.
    However, there are self collisions for $6,448$ ring members, as 3,112 distinct outputs are identified as the true spend of multiple input rings. This amounts to a self collision rate $SCR$ of $0.022$\%, which is likely due to a larger client diversity. This means the actual precision of this heuristic likely isn't this high.
    
    \item To provide another data point on the precision of the heuristic, we also evaluate on the Monero Testnet, where a large dataset of ground-truth transactions exist. There are \nACKJMoneroTestnetKeyImages input rings, of which \nACKJMoneroTestnetKeyImagesSingleTenBlock have exactly one 10-block old ring member. In all but two input rings, the true spend is the 10 block old ring member, yielding a precision of \nACKJMoneroHeuristicPrecision.
    
\end{enumerate}

From the point of the limited ground truth data, the 10 block decoy bug heuristic appears to be highly accurate. Nevertheless, it yields a large, high confidence label set with results between 2018 and 2023, and can be used to assess the quality of other heuristics that are applicable during that time frame. We further discuss these results in Section~\ref{sec:discussion} and proceed with the Differ By One heuristic in the next subsection.
\subsection{Differ By One}
The Differ-by-One heuristic identifies true spends from pairs of input rings characterized by an identical set of ring members, with the exception of a single element. It was previously referenced in a GitHub repository by Kraviec-Thawyer~\cite{krawiec2022ringxor} at the end of 2022, who also provided an implementation to find such instances.

In this heuristic, the distinct ring member is presumed to represent the true spend for that particular input ring. A possible explanation for this pattern is the potential caching of ring members by certain wallet applications, leading to repetitive usage of the same decoys. As this is a true-spend identification heuristic, if the $pk \in R$ identified as the true spend appears in other input rings, it is consequently treated as a decoy in those rings.
The Differ By One pattern has been observed across \nDifferByOneTransactions transactions ranging from \nDifferByOneStartDate until the end of our dataset on \nDifferByOneEndDate.

\begin{figure}[t]
\centering
\includegraphics[width=\linewidth]{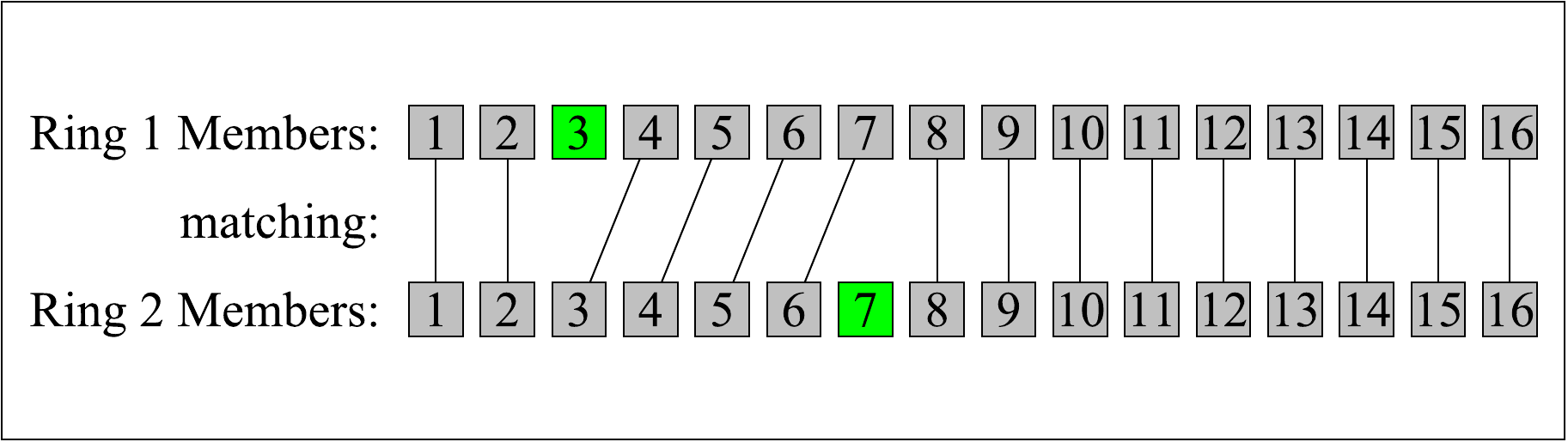}
\caption{Illustration of the Differ-by-One heuristic: given two input rings that are almost identical except for one ring member (i.e. all other ring members match between the rings), the differing outputs (marked in green) are likely the true spends.}
\label{fig:differ-by-one}
\end{figure}

\noindent\textbf{Heuristic Definition.} For every input ring \( R_1 \), if there exists exactly one input ring \( R_2 \) that is almost identical to \( R_1 \) but differs by exactly one ring member, that unique member is hypothesized as the true spend. See Figure~\ref{fig:differ-by-one} for a graphical illustration.

\noindent\textbf{Evaluation.} The heuristic tags $360,102$ ring members as true spent outputs, and a total of $4,777,246$ labeled ring members. For $178$ labeled ring members there exists a self collision, as $89$ outputs are identified as the true spend in multiple input rings, leading to a self collision rate $SCR$ of $0.0037$\%. To validate the heuristic, we treat the zero-mixin, chain reaction and 10 block decoy bug heuristics results applied to Monero mainnet transactions as ground truth. By comparing the Differ-by-One results to this ground truth, we identify $15,729$ incorrectly labeled true spends, and $103,259$ correct ones. By evaluating all labeled ring members, we identify $46,765$ false positives and $460,320$ true positives, yielding a precision of $90.78$\%.
\subsection{Mordinals}

\begin{figure}[h]
\centering
\includegraphics[width=\linewidth]{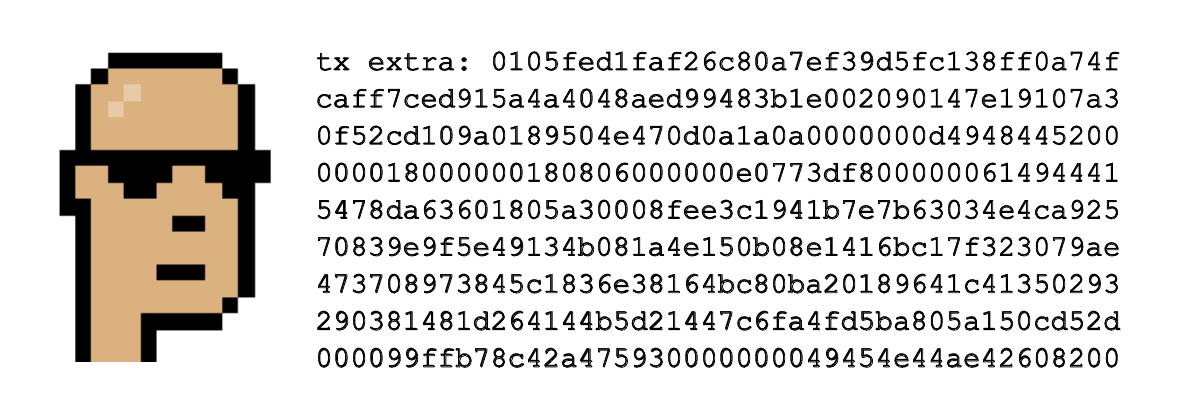}
\caption{Mordinal: An image embedded in the tx\_extra field in transaction hash baa3f1fa73942366c19471aac73b78dd2664eefe634bdbd260d58d09d2a0e259}
\label{fig:mordinal-example}
\end{figure}

In March 2023, the Mordinals protocol was proposed, enabling users to utilize \glspl{nft} on the Monero chain by storing image data in the transaction extra field (tx extra). See Figure~\ref{fig:mordinal-example} for an example. Usage of the protocol was enabled through a forked Monero CLI client that introduced new commands to mint and transfer Mordinals. Identifying Mordinal transactions is possible by parsing the tx extra field for specific mordinal tags, with "\texttt{10}" denoting minting and "\texttt{11}" denoting transfers. The first output in such transactions represents the Mordinal. Mordinal transfers are also distinguishable through their associated decoys; when 15 out of 16 outputs referenced by an input ring are special burned outputs (``\texttt{00000000000000000000000000000000000000000\\00000000000000000000000}'' or ``\texttt{deadbeefdeadbee\\fdeadbeefdeadbeefdeadbeefdeadbeefdeadbeefd\\ead000f}''), the remaining output indicates a Mordinal being transferred.

The burned outputs are non-spendable and therefore obviously decoys if included in any input ring. Mordinal outputs in turn are spendable, but are anticipated to be spent in Mordinal transfer transactions, they can therefore with high likelihood be discarded as decoys when referenced by regular non-Mordinal transactions.

\noindent\textbf{Heuristic Definition.} Define \( MT \) as the set of all Mordinal Minting and Transferring transactions. The Mordinal Decoy Identification Heuristic labels a ring member \( pk_m \) in the input ring \( R \) of any transaction \( tx \) as a decoy if and only if \( tx \) is not part of \( MT \) (i.e., \( tx \notin MT \)) and \( pk_m \) represents the first output of a Mordinal transaction \( tx_m \in MT \).

\noindent\textbf{Evaluation.}
Between 2023-03-09 and 2023-04-21, a total of \nMordinalsMinted Mordinals were minted. Subsequent to this period, transfer transactions have been sparse, and there have been no instances of minting. The heuristic marks \nMordinalDecoys ring members as decoys, of which \nMordinalTruePositives are true positives, and \nMordinalFalsePositives are false positives, when comparing against the heuristics zero mixins, chain reaction and 10 block decoy bug. Overall, the heuristic's precision is therefore \nMordinalPrecision.

\subsection{Coinbase Outputs}
A coinbase transaction is a transaction where a block reward is distributed to miners. There are three types of recipients for coinbase transactions; solo miners, centralized mining pools, and decentralized mining pools, with the latter prominently represented by a service called P2Pool. Unless a solo miner is equipped with significant hardware resources to generate a high mining hashrate, it is beneficial for most miners to join a mining pool in which they are rewarded proportionally to their contributed hashrate for finding a block.

\begin{figure}[!t]
\centering
\includegraphics[width=\linewidth]{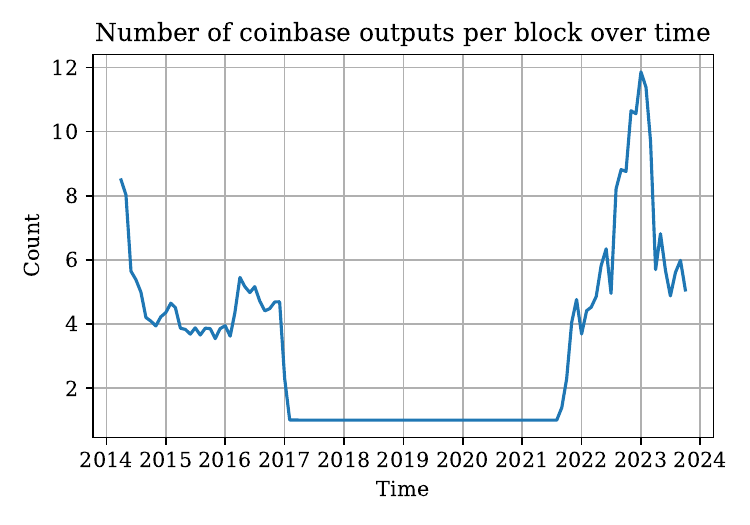}
\caption{The number of coinbase outputs was high until 2017, as Monero used to generate outputs of multiple denominations prior to the introduction of RingCT, hiding amounts. In 2021, the number of outputs started increasing again with the emergence of decentralized mining pool P2Pool.}
\label{fig:coinbase-outputs-per-block-over-time}
\end{figure}

Coinbase transactions associated with centralized pools usually have one output, while those associated with P2Pool usually have a relatively high number of outputs because multiple miners receive payouts in the same transaction. As a consequence, most coinbase outputs are either spent by centralized pools to send payouts to their miners, or spent directly by miners using a decentralized mining pool. Figure \ref{fig:coinbase-outputs-per-block-over-time} shows a spike in the number of coinbase outputs in October 2021 due to the launch of P2Pool, indicating it is responsible for most coinbase outputs. However, the average number of coinbase outputs decreased in March 2023 when P2Pool launched an upgrade that reduces the number of payouts.

Non-miner Monero users are not expected to spend coinbase outputs but the decoy selection algorithm can include them in input rings. As miners tend to receive mining outputs on a recurring basis, they often need to merge their outputs. It is therefore common for coinbase outputs to be spent in transactions that have a relatively high number of inputs. This means that for most small transactions, we should be able to discard referenced coinbase outputs as decoys. However, it is possible for miners to merge their outputs in small transactions, so we experiment with multiple thresholds for the maximum number of inputs of transactions for which we can discard referenced coinbase outputs. 

\begin{figure}[!t]
\centering
\includegraphics[width=\linewidth]{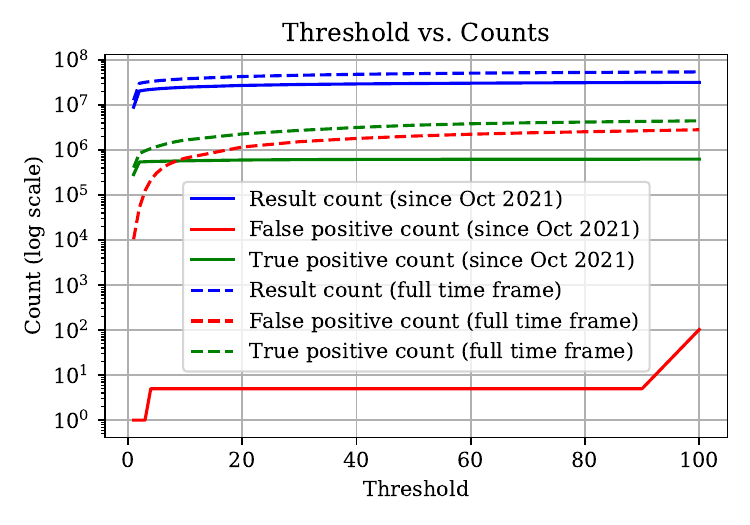}
\caption{If we apply the heuristic to all transactions, both the number of false positives and true positives would be high. However, if we only consider transactions that have happened since P2Pool was launched in October 2021, the number of true positives remains high, and the number of false positives becomes low for transactions that have $\leq 90$ inputs.}
\label{fig:coinbase-threshold-metrics}
\end{figure}

Figure \ref{fig:coinbase-threshold-metrics} shows how the number of excluded decoys, false positives, and true positives change as the threshold increases. If we apply the heuristic to all transactions, both the number of false positives and true positives would be high. However, if we only consider transactions that have happened since P2Pool was launched in October 2021, the number of true positives remains high, and the numbers of false positives becomes low for all thresholds that are $\leq 90$. Therefore, we apply the heuristic to all transactions that have happened since October 2021, for which the number of input rings is $\leq 90$.

\noindent\textbf{Heuristic Definition.} Define \( C \) as the set of all coinbase transactions. The heuristic labels a ring member \( pk \) in the input ring \( R \) of any transaction \( tx \) as a decoy if and only if the number of inputs of \( tx \) is $\leq 90$, the date of \( tx \) is $\geq$ \mbox{2021-10-01} and \( pk \) is an output of a coinbase transaction \( tx_c \in C \).

\noindent\textbf{Evaluation.} The heuristic marks $31,650,837$ ring members as decoys, of which $626,679$ are true positives, and $5$ are false positives, when comparing against the heuristics zero mixins, chain reaction and 10 block decoy bug. The heuristic's precision is therefore 99.9\%.

\begin{table*}[!ht]
\caption{Comparing heuristics: pairwise collision and agreement rates between the heuristics. Collision rates are mostly very low.\\Only coinbase and P2Pool have a higher agreement rate, meaning most other heuristics are very complementary to each other. Ideally, heuristics have a bit of agreement, and very low collision rates.
}
\centering
\renewcommand{\arraystretch}{1.5}
\begin{tabular}{@{}rccccccc}
\toprule
& 0-Mix + Chain React. & Differ By One & 10 Block Decoy & Coinbase & Mordinals & P2Pool Out. Merging & \\
\midrule
0-Mix + Chain React. & \cellcolor{gray}\color{white}\nSelfCollisionsZeroMixCascade \nOverlapChainReaction & \nAgreementsDifferByOneZeroMixCascade & \nAgreementsTenBlockZeroMixCascade & \nAgreementsCoinbaseZeroMixCascade & \nAgreementsMordinalsZeroMixCascade & \nAgreementsPPoolZeroMixCascade & \multirow{5}{*}{\rotatebox[origin=c]{90}{Agreement Rate}}\\
Differ By One & \cellcolor{gray}\color{white}\nCollisionsDifferByOneZeroMixCascade \nOverlapChainReactionDifferByOne & \cellcolor{gray}\color{white}\nSelfCollisionsDifferByOne \nOverlapDifferByOne & \nAgreementsTenBlockDifferByOne & \nAgreementsCoinbaseDifferByOne & \nAgreementsMordinalsDifferByOne & \nAgreementsPPoolDifferByOne & \\
10 Block Decoy & \cellcolor{gray}\color{white}\nCollisionsTenBlockZeroMixCascade \nOverlapChainReactionTenBlock & \cellcolor{gray}\color{white}\nCollisionsTenBlockDifferByOne \nOverlapDifferByOneTenBlock & \cellcolor{gray}\color{white}\nSelfCollisionsTenBlock \nOverlapTenBlock & \nAgreementsCoinbaseTenBlock & \nAgreementsMordinalsTenBlock  & \nAgreementsPPoolTenBlock & \\
Coinbase & \cellcolor{gray}\color{white}\nCollisionsCoinbaseZeroMixCascade \nOverlapChainReactionCoinbase & \cellcolor{gray}\color{white}\nCollisionsCoinbaseDifferByOne \nOverlapDifferByOneCoinbase & \cellcolor{gray}\color{white}\nCollisionsCoinbaseTenBlock \nOverlapTenBlockCoinbase & \cellcolor{gray}\color{white}\nSelfCollisionsCoinbase \nOverlapCoinbase & \nAgreementsMordinalsCoinbase  & \nAgreementsPPoolCoinbase & \\
Mordinals & \cellcolor{gray}\color{white}\nCollisionsMordinalsZeroMixCascade \nOverlapChainReactionMordinals & \cellcolor{gray}\color{white}\nCollisionsMordinalsDifferByOne \nOverlapDifferByOneMordinals & \cellcolor{gray}\color{white}\nCollisionsMordinalsTenBlock \nOverlapTenBlockMordinals & \cellcolor{gray}\color{white} \nCollisionsMordinalsCoinbase \nOverlapCoinbaseMordinals & \cellcolor{gray}\color{white}\nSelfCollisionsMordinals \nOverlapMordinals  & \nAgreementsPPoolMordinals & \\
P2Pool Out. Merging & \cellcolor{gray}\color{white}\nCollisionsPPoolZeroMixCascade \nOverlapChainReactionPPool & \cellcolor{gray}\color{white}\nCollisionsPPoolDifferByOne \nOverlapDifferByOnePPool & \cellcolor{gray}\color{white}\nCollisionsPPoolTenBlock \nOverlapTenBlockPPool & \cellcolor{gray}\color{white} \nCollisionsPPoolCoinbase \nOverlapCoinbasePPool & \cellcolor{gray}\color{white}\nCollisionsPPoolMordinals \nOverlapMordinalsPPool & \cellcolor{gray}\color{white}\nSelfCollisionsPPool \nOverlapPPool & \\
& \multicolumn{6}{c}{\cellcolor{gray}\color{white}Collision Rate} & \cellcolor{gray}\\
\bottomrule
\end{tabular}
\label{tab:heuristic-agreements-collisions2}
\end{table*}

\subsection{P2Pool Output Merging}
\label{subsec:p2pool_output_merging}
\begin{figure}[b]
\centering
\includegraphics[width=\linewidth]{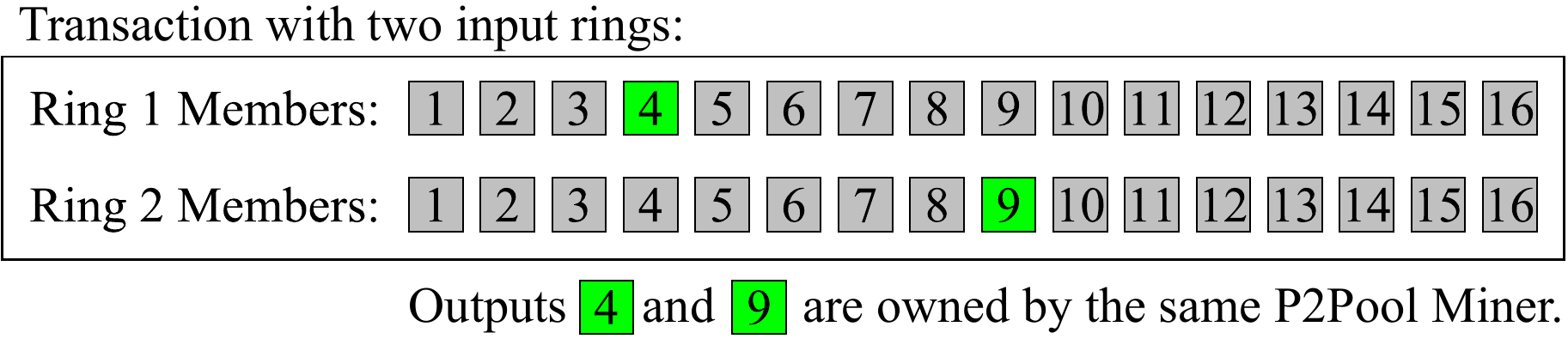}
\caption{Example of P2Pool output merging transaction: P2Pool states output ownership (green indicates same owner) for their coinbase transactions. Therefore, the output merging heuristic can be applied on the outputs of the same miner, identifying those outputs as true spends.}
\label{fig:p2pool_output_merging}
\end{figure}

\begin{figure}[h]
\centering
\vspace{3em}
\includegraphics[width=0.98\linewidth]{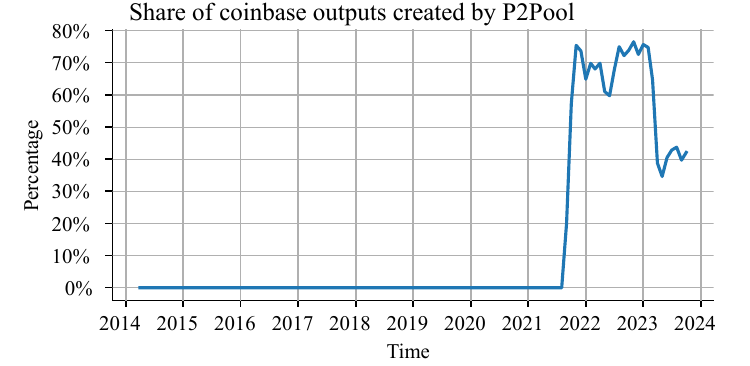}
\caption{The share of coinbase outputs created by P2Pool in comparison to all generated coinbase outputs spiked to more than $70\%$ at the end of 2021, and still accounts for about $40\%$ near the end of 2023.}
\label{fig:share-p2pool-outputs}
\end{figure}

Many mining pools post their transactions online, often including the associated addresses of the miners. Transactions that reference multiple outputs from multiple known mining pool transactions are likely generated either by a miner consolidating their payouts or by a mining pool merging its change outputs for subsequent payouts.
Since P2Pool owns a significant share of coinbase outputs (see Figure~\ref{fig:share-p2pool-outputs}) and they publish all transaction hashes and miner addresses, we apply this heuristic to their outputs. However, the analysis can be extended to centralized mining pools as well ~\cite{wijaya2021}. We consider transactions where all rings reference outputs owned by the same miner. See Figure~\ref{fig:p2pool_output_merging} for an example. In some cases, a single output is referenced by multiple transactions. For those outputs, we pick the transaction that has the highest number of referenced outputs owned by a single miner.

\noindent\textbf{Heuristic Definition.} Define $M$ as the set of all P2Pool miners and $O_m$ as the set of outputs owned by a miner \(m \in M\). For every \(o \in O_m\), consider the set of transactions $T_o$ for which there exists at least one $pk$ for every ring member $R$ such that \(pk \in R\) and \(pk \in O_m\). If $T_o$ has multiple transactions, we only consider the one with the highest number of referenced outputs owned by $m$. $pk$ is likely to be the true spend of $R$.

\noindent\textbf{Evaluation.} The heuristic tags $11,368$ ring members as true spent outputs, and a total of $269,124$ labeled ring members. For $99$ labeled ring members there exists a self collision, as $2$ outputs are identified as the true spend in multiple input rings, and $48$ input rings have multiple ring members identified by the heuristic as the true spend. This leads to a self collision rate $SCR$ of $0.037$\%. To validate the heuristic, we treat the zero-mixin, chain reaction and 10 block decoy bug heuristics results applied to Monero mainnet transactions as ground truth. By comparing the results to this ground truth, we identify an overlap of $142$ spent outputs, all of which are labeled incorrectly by the heuristic. However, by comparing all labeled ring members, we identify $284$ false positives and $4,963$ true positives, yielding a precision of $94.59$\%.

An attempt to improve performance of this heuristic would be requiring more outputs to be merged in single transactions at the cost of identifying fewer output merging transactions and true spends.

\begin{figure*}[!t]
\centering
\includegraphics[width=\linewidth]{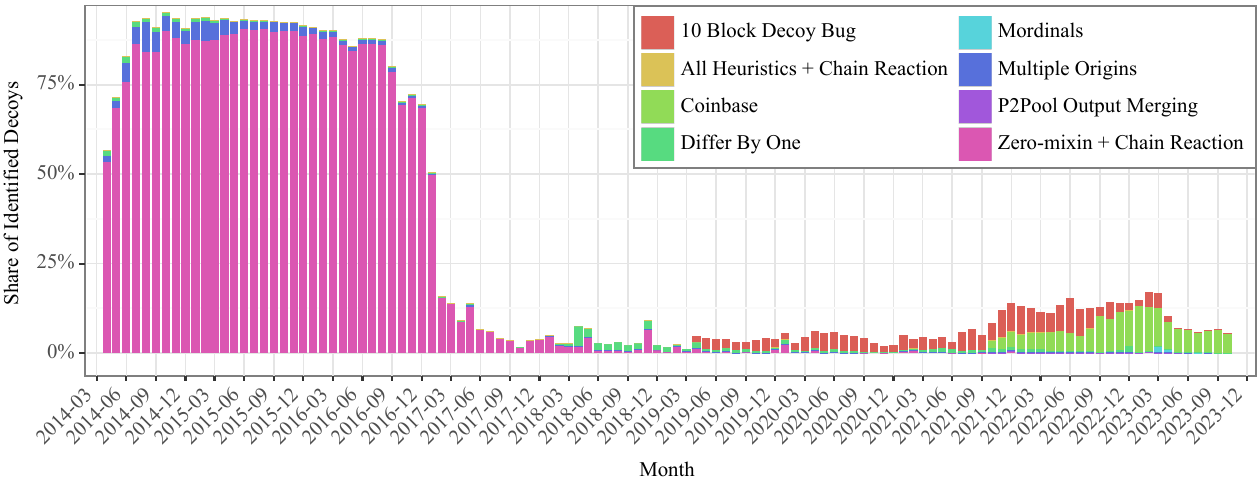}
\caption{Share of Identified Decoys per month, colored by heuristic. In alignment with earlier works, the Zero-mixin + Chain Reaction heuristic was very effective between 2014 and 2018. Differ By One has an overlap during that time pointing to this particular bug primarily being present in the early years of Monero. Most interestingly, the 10 Block Decoy Bug heuristic is among the most effective in recent years, abruptly ending with the vulnerability disclosure in May of 2023. Of all heuristics, the coinbase decoy identification heuristic remains the most applicable at the end of 2023.}
\label{fig:decoy_share_over_time}
\end{figure*}

\subsection{Combined Analysis}
When evaluating heuristics in the previous sections, we focused on the precision. We now turn to a pairwise comparison of the proposed heuristics. Given heuristics $H_1$ and $H_2$ that each make a statement about a ring member $pk$ in $R$, an agreement means both heuristics yield the same label, and a collision otherwise. This has the benefit that we can compare the results of a true-spend identification heuristic with a decoy identification heuristic.

Let $|H_1|$ and $|H_2|$ be the number of ring members labeled by heuristics $H_1$ and $H_2$ respectively. We denote the number of agreements between two heuristics as $A$ and the number of collisions as $C$. This is similar to $TP$ and $FP$, but we intentionally use a different notation as we technically do not have ground truth available for all pairwise comparisons.

\noindent We measure collision rate and agreement rate that we define as follows:
\[Collision\ Rate = \frac{C}{A+C}\]
\[Agreement\ Rate = \frac{A}{min(|H_1|,|H_2|)}\]

\begin{figure}[!t]
\centering
\includegraphics[width=\linewidth]{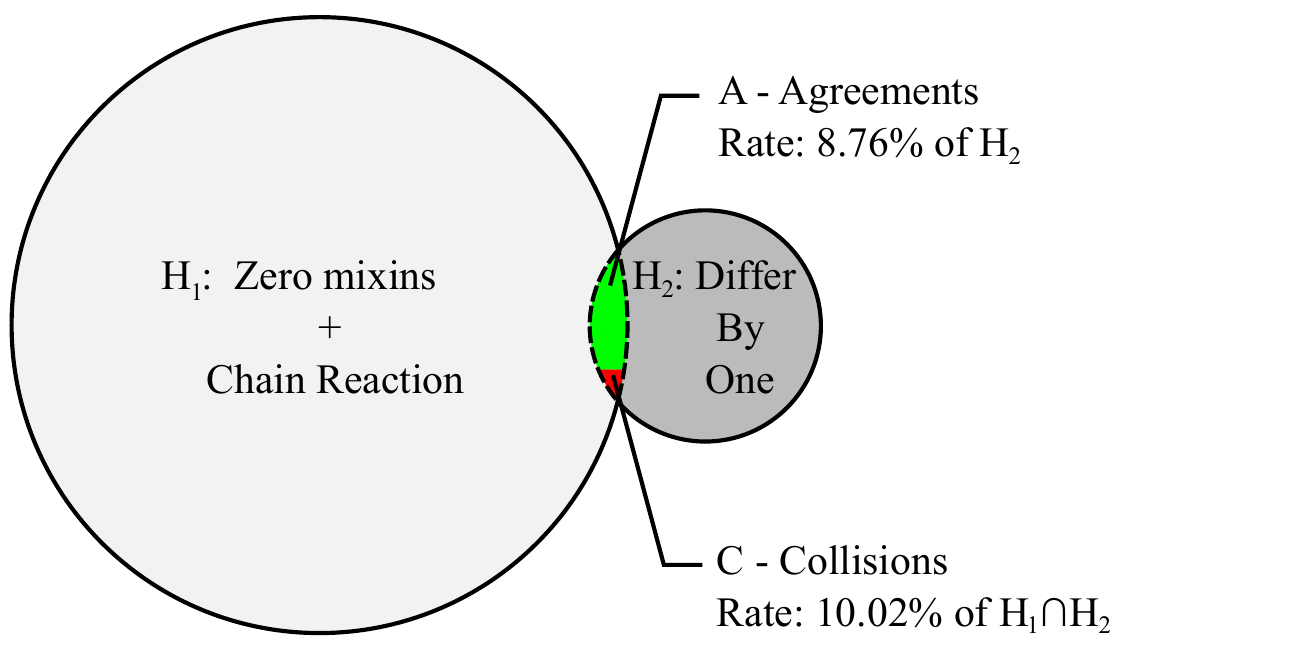}
\caption{Agreements, Collisions and their rates illustrated with a Venn diagramm using heuristics Zero mixins + Chain reaction and Differ By One. In the intersection, both heuristics make statements about the same ring members.}
\label{fig:comparison-zero-mixins-differ-by-one}
\end{figure}

To illustrate collision and agreement rate, consider Figure~\ref{fig:comparison-zero-mixins-differ-by-one}, which compares the Differ By One Heuristic with Zero-Mixins and Chain Reaction in a Venn diagram. The size of each heuristic's area corresponds to the number of ring member labels each generates. In the intersection, both heuristics make statements about the same ring members. $10.02\%$ of the intersection are collisions, i.e. conflicting statements. The agreements in comparison to the smaller heuristic is the agreement rate, in this case $8.76\%$. A very high agreement rate would therefore mean that the heuristic is mostly contained in another heuristic, and therefore lacks novelty. A great heuristic would exhibit the following properties: some agreement with other heuristics for validation, very low collision rate and a high number of ring members it makes a statement about.

Table \ref{tab:heuristic-agreements-collisions2} shows that collision rates are mostly low, with a couple of exceptions. Differ By One has a collision rate of 10\% with Zero-mixin + Chain Reaction, which we have already seen in the heuristic evaluation section. Coinbase also has a collision rate of 10.18\% with P2Pool Output Merging. This points to the precision of the Coinbase heuristic being lower than previously estimated. We believe this heuristic is more useful when used with additional context. For example, it is safer to use if we have additional information that confirms the transaction of interest was not made by a miner or a mining pool. The agreement rate is low for most heuristics, which indicates that the heuristics are complementary to each other. The highest agreement of 36.42\% exists between the Coinbase heuristic and the P2Pool Output Merging heuristics. This is expected since the P2Pool Output Merging heuristic leads to coinbase outputs being marked as decoys elsewhere.

Figure \ref{fig:decoy_share_over_time} shows the percentage of decoys identified by each heuristic over time. Zero-mixin + Chain Reaction can be used effectively to evaluate the impact of heuristics for old transactions, but they have almost no impact on recent transactions. For recent transactions, 10 Block Decoy Bug and the Coinbase heuristic have the highest impact.

By applying the chain reaction heuristic to the results of all previously described heuristics, combined with the results of the zero-mixin heuristic, we identify \nChainReactionIncrementalSpentOutputs additional true spent outputs and \nChainReactionIncrementalKnownDecoys additional decoys. The impact is low compared to other heuristics, but it is worth noting that those spent outputs and decoys can only be identified by combining results of all heuristics.

\begin{figure}[!t]
\centering
\includegraphics[width=\linewidth]{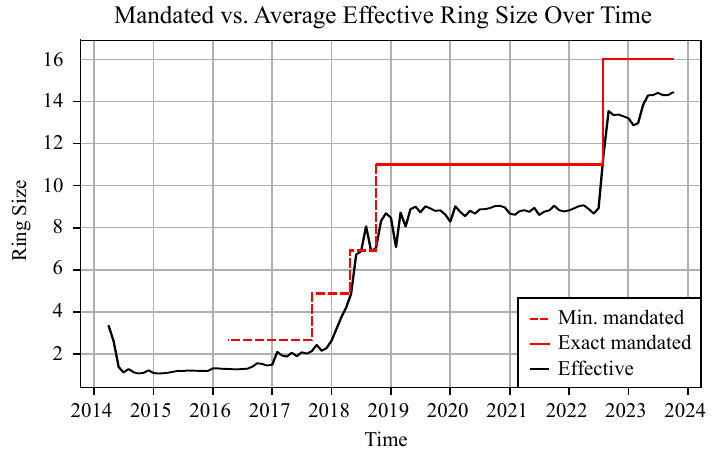}
\caption{Throughout the past years, the mandatory ring size of Monero input rings was increased with protocol upgrades. Between 2018 and 2022, the mandated ring size was 11, and we can show that the average effective ring size has been around 9. With the latest upgrade in August 2022, the mandatory ring size was increased to 16. After applying the heuristics described in this paper, the average effective ring size is still above 14 in October of 2023.}
\label{fig:avg-effective-ring-size}
\end{figure}

Finally, Figure \ref{fig:avg-effective-ring-size} shows how the effective ring size has changed over time. In Augest 2022, the mandatory ring size was increased to 16, but the effective ring size was lower than 14, primarily because of the spike in the number of coinbase outputs caused by P2Pool. The effective ring size started to go up in early 2023 after P2Pool introduced an upgrade that reduces the number of payouts and thereby reduces the number of coinbase outputs they generated. Overall, the figure shows that the heuristics described in this paper have some impact on the effective ring size, but the impact is relatively low since the effective ring size is still higher than 14.

\section{Discussion}
\label{sec:discussion}
We now turn to discussing the results of our heuristic analyses. In general, for most heuristics the collision rate is low. There exists some agreement between several heuristics, but ultimately the size of the ground truth originating from the zero-mixin and chain reaction heuristics is very small since 2018, and will diminish even further. The 10 block decoy bug heuristic can therefore be considered as the best alternative among the available options, which is the reason we've used it to evaluate some of the other heuristics. Nevertheless, the 10 block decoy bug heuristic only works really well on input rings generated by the wallet2 library. In reality, its precision is likely slightly lower than what we were able to determine, as there is more client diversity on the Monero mainnet.

The Monero testnet dataset could not be used to evaluate all of the heuristics, as it does not contain transactions spending Mordinals or coinbase outputs, and does not contain outputs from P2Pool, or transactions that exhibit the Differ By One phenomenon. Regarding the latter heuristic, we stipulate that the origin of this pattern is that there exists one or more wallet applications that do not sample ring members correctly, and instead perhaps cache a list of previously used decoys.
To mitigate the impact of the described heuristics on Monero users' privacy, wallet and service operators should check for these bugs, and users should use the latest software versions.

We did not include an analysis of the existing heuristics closed set and intersection sets originating from Monero forks as we wanted to focus on recent developments instead.
We could have used the ground truth result of the dulmage-mendelsohn decomposition as proposed by Vijayakumaran~\cite{vijayakumaran2023analysis}, but
the results are nearly identical to using zero mixins + chain reaction, so we opted for lower implementation complexity over implementing the underlying algorithm.

Apart from the Differ By One heuristic, in particular the P2Pool Output Merging heuristic has had higher collisions with the Coinbase exclusion heuristic. With the popularization of P2Pool, the Coinbase heuristic is not always correct. It is common for miners to spend their outputs in transactions that have less than 90 input rings, and those outputs would be falsely marked as decoys by the Coinbase heuristic. However, as a result of the spike in the number of coinbase outputs that was caused by P2Pool, most coinbase ring members are actually decoys, which means the heuristic is correct more often than not. Monero already has a proposal to avoid selecting coinbase outputs as decoys\footnote{\url{https://github.com/monero-project/research-lab/issues/109}}. We recommend applying the results of those heuristics very cautiously. The Coinbase heuristic in particular is valid in isolated scenarios where additional context is available, such as knowledge that the transaction sender is not a miner.
This view is reinforced by Deuber et al.~\cite{deuber2022sok}, who show that deanonymization heuristics for cryptocurrencies are assumption-based and prone to false positives. Therefore, it's critical to use these heuristics cautiously, acknowledging the possibility of errors.

\section{Conclusion}
This paper contributes significant new insights to the field of cryptocurrency privacy, particularly within the Monero ecosystem, focusing on developments between 2019 and 2023. Our comprehensive work delves into the intricacies of several key heuristics, including the '10 Block Decoy Bug', 'Differ-by-one', Coinbase, Mordinal decoy identification, and a P2Pool specific output merging heuristic, grounded in the analysis of publicly known miner payouts. While these topics have been discussed within the Monero community on platforms like Github and Reddit, our study stands out by providing a systematic, comparative, and combined evaluation of these methodologies.

Our findings illustrate that most of these heuristics demonstrate high precision, with the '10 Block Decoy Bug' and the Coinbase decoy identification heuristics having the most significant impact in the period from 2019 to 2023. Notably, the '10 Block Decoy Bug' heuristic can serve as an evaluation baseline for future heuristics applicable within this timeframe. 

A crucial aspect of our analysis is the measurement of each heuristic's impact over time, including the assessment of the effective ring size of Monero's transaction inputs up until October 2023. This comprehensive approach has enabled us to provide a nuanced and detailed understanding of the current state of privacy in the Monero ecosystem. 

This work can be extended in the future by applying the output merging heuristic to transactions associated with centralized mining pools. We can also evaluate two versions of the coinbase heuristic separately: one that is applied only to P2Pool outputs and another that is applied to centralized mining pool outputs.

Finally, we want to highlight that every heuristic is based on certain assumptions and can yield false positives. This underlines the necessity of acknowledging the potential for inaccuracies and exercising caution when applying these heuristics in real-world scenarios. Our findings contribute not only to the academic discourse but also offer practical insights that could guide future developments in studying privacy in cryptocurrency transactions.

\section*{Acknowledgment}
We extend our sincere thanks to Justin Ehrenhofer and Bernhard Haslhofer for their valuable feedback and insights on earlier drafts of this manuscript. Their expertise and detailed reviews greatly enhanced the paper's quality and clarity. Lastly, we are grateful for the anonymous peer reviews that informed the final version of this paper.

\bibliographystyle{IEEEtran}
\bibliography{references}

\begin{thebibliography}{10}
\providecommand{\url}[1]{#1}
\csname url@samestyle\endcsname
\providecommand{\newblock}{\relax}
\providecommand{\bibinfo}[2]{#2}
\providecommand{\BIBentrySTDinterwordspacing}{\spaceskip=0pt\relax}
\providecommand{\BIBentryALTinterwordstretchfactor}{4}
\providecommand{\BIBentryALTinterwordspacing}{\spaceskip=\fontdimen2\font plus
\BIBentryALTinterwordstretchfactor\fontdimen3\font minus
  \fontdimen4\font\relax}
\providecommand{\BIBforeignlanguage}[2]{{%
\expandafter\ifx\csname l@#1\endcsname\relax
\typeout{** WARNING: IEEEtran.bst: No hyphenation pattern has been}%
\typeout{** loaded for the language `#1'. Using the pattern for}%
\typeout{** the default language instead.}%
\else
\language=\csname l@#1\endcsname
\fi
#2}}
\providecommand{\BIBdecl}{\relax}
\BIBdecl

\bibitem{van2013cryptonote}
N.~van Saberhagen, ``Cryptonote v 2. 0 (white paper),'' 2013.

\bibitem{noether2016ring}
S.~Noether, A.~Mackenzie \emph{et~al.}, ``Ring confidential transactions,''
  \emph{Ledger}, vol.~1, pp. 1--18, 2016.

\bibitem{deuber2022sok}
D.~Deuber, V.~Ronge, and C.~R{\"u}ckert, ``Sok: Assumptions underlying
  cryptocurrency deanonymizations,'' \emph{Proceedings on Privacy Enhancing
  Technologies}, vol.~3, pp. 670--691, 2022.

\bibitem{ghesmati2022sok}
S.~Ghesmati, W.~Fdhila, and E.~Weippl, ``Sok: How private is bitcoin?
  classification and evaluation of bitcoin privacy techniques,'' in
  \emph{Proceedings of the 17th International Conference on Availability,
  Reliability and Security}, 2022, pp. 1--14.

\bibitem{victor2020address}
F.~Victor, ``Address clustering heuristics for ethereum,'' in \emph{Financial
  Cryptography and Data Security: 24th International Conference, FC 2020, Kota
  Kinabalu, Malaysia, February 10--14, 2020 Revised Selected Papers 24}.\hskip
  1em plus 0.5em minus 0.4em\relax Springer, 2020, pp. 617--633.

\bibitem{kumar2017traceability}
A.~Kumar, C.~Fischer, S.~Tople, and P.~Saxena, ``A traceability analysis of
  monero's blockchain,'' in \emph{Computer Security -- ESORICS 2017}, S.~N.
  Foley, D.~Gollmann, and E.~Snekkenes, Eds.\hskip 1em plus 0.5em minus
  0.4em\relax Cham: Springer International Publishing, 2017, pp. 153--173.

\bibitem{moser2018empirical}
M.~M{\"o}ser, K.~Soska, E.~Heilman, K.~Lee, H.~Heffan, S.~Srivastava, K.~Hogan,
  J.~Hennessey, A.~Miller, A.~Narayanan \emph{et~al.}, ``An empirical analysis
  of traceability in the monero blockchain,'' \emph{Proceedings on Privacy
  Enhancing Technologies}, vol. 2018, no.~3, 2018.

\bibitem{hinteregger2019short}
A.~Hinteregger and B.~Haslhofer, ``Short paper: An empirical analysis of monero
  cross-chain traceability,'' in \emph{Financial Cryptography and Data
  Security: 23rd International Conference, FC 2019, Frigate Bay, St. Kitts and
  Nevis, February 18--22, 2019, Revised Selected Papers 23}.\hskip 1em plus
  0.5em minus 0.4em\relax Springer, 2019, pp. 150--157.

\bibitem{yu2019new}
Z.~Yu, M.~H. Au, J.~Yu, R.~Yang, Q.~Xu, and W.~F. Lau, ``New empirical
  traceability analysis of cryptonote-style blockchains,'' in
  \emph{International Conference on Financial Cryptography and Data
  Security}.\hskip 1em plus 0.5em minus 0.4em\relax Springer, 2019, pp.
  133--149.

\bibitem{nakamoto2008bitcoin}
S.~Nakamoto, ``Bitcoin: A peer-to-peer electronic cash system,'' 2008.

\bibitem{bunz2018bulletproofs}
B.~B{\"u}nz, J.~Bootle, D.~Boneh, A.~Poelstra, P.~Wuille, and G.~Maxwell,
  ``Bulletproofs: Short proofs for confidential transactions and more,'' in
  \emph{2018 IEEE symposium on security and privacy (SP)}.\hskip 1em plus 0.5em
  minus 0.4em\relax IEEE, 2018, pp. 315--334.

\bibitem{wijaya2021transparency}
D.~A. Wijaya, J.~K. Liu, R.~Steinfeld, and D.~Liu, ``Transparency or anonymity
  leak: Monero mining pools data publication,'' in \emph{Information Security
  and Privacy: 26th Australasian Conference, ACISP 2021, Virtual Event,
  December 1--3, 2021, Proceedings 26}.\hskip 1em plus 0.5em minus 0.4em\relax
  Springer, 2021, pp. 433--450.

\bibitem{wijaya2019unforkability}
D.~A. Wijaya, J.~K. Liu, R.~Steinfeld, D.~Liu, and J.~Yu, ``On the
  unforkability of monero,'' in \emph{Proceedings of the 2019 ACM Asia
  Conference on Computer and Communications Security}, 2019, pp. 621--632.

\bibitem{vijayakumaran2023analysis}
\BIBentryALTinterwordspacing
S.~Vijayakumaran, ``{Analysis of CryptoNote Transaction Graphs Using the
  Dulmage-Mendelsohn Decomposition},'' in \emph{5th Conference on Advances in
  Financial Technologies (AFT 2023)}, ser. Leibniz International Proceedings in
  Informatics (LIPIcs), J.~Bonneau and S.~M. Weinberg, Eds., vol. 282.\hskip
  1em plus 0.5em minus 0.4em\relax Dagstuhl, Germany: Schloss Dagstuhl --
  Leibniz-Zentrum f{\"u}r Informatik, 2023, pp. 28:1--28:22. [Online].
  Available:
  \url{https://drops.dagstuhl.de/entities/document/10.4230/LIPIcs.AFT.2023.28}
\BIBentrySTDinterwordspacing

\bibitem{dulmagemendelsohn1958}
A.~L. Dulmage and N.~S. Mendelsohn, ``Coverings of bipartite graphs,''
  \emph{Canadian Journal of Mathematics}, vol.~10, p. 517–534, 1958.

\bibitem{aeeneh2021new}
S.~Aeeneh, J.~O. Chervinski, J.~Yu, and N.~Zlatanov, ``New attacks on the
  untraceability of transactions in cryptonote-style blockchains,'' in
  \emph{2021 IEEE International Conference on Blockchain and Cryptocurrency
  (ICBC)}.\hskip 1em plus 0.5em minus 0.4em\relax IEEE, 2021, pp. 1--5.

\bibitem{rucknium2023reddit}
{u/Rucknium}, ``Empirical privacy impact of mordinals (monero nfts),''
  \url{https://www.reddit.com/r/Monero/comments/12kv5m0/empirical_privacy_impact_of_mordinals_monero_nfts/},
  April 2023, reddit post on r/Monero, accessed September 12, 2023.

\bibitem{ackj2022lotr}
ACK-J, ``Lord of the rings: An empirical analysis of monero's ring signature
  resilience to artificially intelligent attacks,'' Monero Research Lab,
  GitHub, Technical Report, 8 2022, final Report for Multidisciplinary Academic
  Grants in Cryptocurrencies. Available online:
  \url{https://raw.githubusercontent.com/ACK-J/Monero-Dataset-Pipeline/main/Lord_of_the_Rings__An_Empirical_Analysis_of_Monero_s_Ring_Signature_Resilience_to_Artificially_Intelligent_Attacks.pdf}.

\bibitem{krawiec2022ringxor}
M.~P. Krawiec-Thayer, ``ringxor,''
  \url{https://github.com/Mitchellpkt/ringxor}, 2022.

\bibitem{wijaya2021}
D.~Wijaya, J.~Liu, R.~Steinfeld, and D.~Liu, \emph{Transparency or Anonymity
  Leak: Monero Mining Pools Data Publication}, 11 2021, pp. 433--450.

\end{thebibliography}

\end{document}